\documentclass[aps,prb,showpacs,showkeys]{revtex4-1}
\usepackage[T2A,OT1]{fontenc}
\usepackage[cp1251]{inputenc}
\usepackage[russian,ukrainian,english]{babel}
\usepackage[dvips]{epsfig}            
\usepackage{latexsym}
\usepackage{amsfonts}
\usepackage{xcolor}[2007/01/21]
\usepackage[dvipdfm,bookmarks=true,bookmarksnumbered,bookmarksopen]{hyperref}
\hypersetup{pdfpagemode=FullScreen,colorlinks=true,
citecolor=green,unicode,bookmarksopenlevel=3,
pdftoolbar=true,pdfauthor={Helen Gomonay},pdftitle={Spin transfer in AFM}}
\usepackage{amsbsy}         

\begin{document}
\title{MAGNETOELASTIC COUPLING AND POSSIBILITY OF SPINTRONIC
ELECTROMAGNETOMECHANICAL EFFECTS}
\author{Helen V. Gomonay}\author{Svitlana V. Kondovych}
\affiliation {
 National Technical University of Ukraine ``KPI''\\ ave Peremogy, 37, 03056, Kyiv,
Ukraine}

\author{Vadim M. Loktev}
\affiliation {Bogolyubov Institute for Theoretical Physics NAS of
Ukraine,\\ Metrologichna str. 14-b, 03143, Kyiv, Ukraine}
\begin{abstract}
Nanoelectromangetomechanical systems (NEMMS) open up a new path for the
development of high speed autonomous nanoresonators and signal
generators that could be used as actuators, for information
processing, as elements of quantum computers etc. Those NEMMS that
include ferromagnetic layers could be controlled by the electric
current due to effects related with spin transfer. In the present
paper we discuss another situation when the current-controlled behaviour of nanorod that
includes an antiferro- (instead of one of ferro-) magnetic  layer.
We argue that in this  case ac spin-polarized current can also induce resonant coupled
magneto-mechanical oscillations and produce an oscillating
magnetization of antiferromagnetic (AFM) layer.  These effects are
caused by \emph{i}) spin-transfer torque exerted to AFM at the
interface with nonmagnetic spacer and by \emph{ii}) the effective
magnetic field produced by the spin-polarized free electrons due to
$sd$-exchange.The described nanorod with an AFM layer can
find an application in magnetometry and as a current-controlled
high-frequency mechanical oscillator.

\end{abstract}
\maketitle

\section{Introduction}
\label{sec_Intro}
 Nanoelectromagnetomechanical systems (NEMMS) that
convert electromagnetic energy into mechanical motion and
\emph{vice versa} are now of great interest for several reasons.
First of all, NEMMS themselves give yet another manifestation of
the coupling between magnetic and mechanical degrees of freedom.
Up to now magneto-mechanical interactions were the most completely
studied for the systems with no electric current (we are talking
about the orientational phase transitions, see, e.g.
\cite{Eremenko_Sirenko(e)}, the coupled magnon-phonon  modes
\cite{Akhiezer:1958}, formation of a magneto-elastic gap
\cite{Borovik-Romanov:1984} etc.). In these cases one can speak
about thermodynamic equilibrium and describe the system with the
time-independent equations. At the same time in recent years
investigations in physics of magnetic phenomena have moved to a
new field  spintronics, where not just the current, but the
\emph{spin-polarized} electrical current is a critical component
that forms the magnetic properties of -- mainly metallic --
systems.

On the other hand, recently increased attention to NEMMS is also
related with their potential applications. In particular, because
of small geometrical size, the fundamental mechanical modes of
NEMMS fall into GHz range and corresponding devices could be used
as high-frequency actuators and transducers of mechanical motion
\cite{Roukes:2007} (see also recent review \cite{Eom:2011PhR} and
references therein). Besides, at low temperatures (much smaller
than the energy of fundamental mode) NEMMS show quantized
mechanical behavior and thus could be used for the quantum
measurements and quantum information processing
\cite{Zhao:2008CoTPh, Savelev2007PhRvB75p5417S,
Cleland:2004PhRvL93g0501C, Garanin:2011arXiv1104.1170G}. At last,
due to high sensitivity to the external fields, including
electric, magnetic and surface stresses, the NEMMS could be used
as the effective tools for biological imaging \cite{Eom:2011PhR},
magnetometry \cite{davis:072513, losby:123910}, for the
measurement of magnetoelastic properties and magnetic anisotropy
of the materials \cite{Masmanidis:2005PhRvL} etc.

An effective way to induce nanomechanical oscillations is based on
the spin-related phenomena, in particular, on spin transferred
torque (STT) predicted by Berger \cite{Berger:1996} and
Slonczewski \cite{SLonczewski:1989, Slonczewski:1996}. Flip of the
free electron spin at the interfaces between the layers with
different magnetic properties is related with the change of the
angular momentum and for nanosize objects (like NEMMS) can result
in the noticeable rotation, torsion or bending of the sample.

Up to dates, combination of nanomechanics and spintronics is
implemented in the devices that include ferromangetic (FM) and
nonmagnetic (NM) metallic layers. In a nanowire with an only
  FM/NM interface the FM layer servers as a polarizer for an electric current, and spin
  flip processes at the FM/NM interface
  produce a mechanical torque  in the sample \cite{fulde-1998-7, Mohanty:2004, Mohanty:PhysRevB.66.085416,
  Mohanty:2008}. Another modification of NEMMS (see \cite{Kovalev:2006, kovalev-2007-75,
  kovalev-2008-101}) is analogous to spin-valves and includes at least two FM
  layers -- one is a polarizer and the magnetization of the other is
  rotated by STT. Oscillations of magnetization, in turn, induce the mechanical movement, due to the presence of spin-lattice
  coupling.

\begin{figure}[htbp]
\begin{centering}
 \includegraphics[width=0.5\columnwidth]{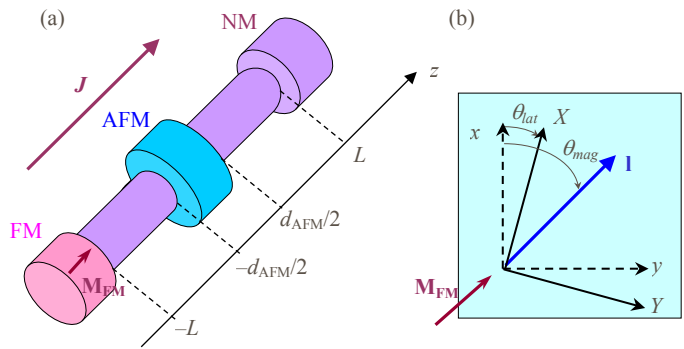}
  \caption{(Color online) \textbf{Nanotorsional oscillator}. Nanorod made
  of NM metal with thin AFM section is mechanically clamped between the
  FM and NM leads (a). The current $J$ that flows
  from FM to NM lead is polarized in $\mathbf{M}_{\mathrm{FM}}\|Z$
  direction and gives rise to the torques twisting the AFM vector $\mathbf{l}$ in
  the middle section (b). Due to magnetic anisotropy, rotation of
  the  magnetic moments through the angle $\theta_\mathrm{mag}$ induces rotation of
  the crystal lattice through the angle $\theta_\mathrm{lat}$. Axes $x$, $y$ denote
the reference frame, while $X$, $Y$ show the instantaneous
orientation of the rotated crystal axes.
 \label{fig_nanorod}}
 \end{centering}
\end{figure}

In the present paper we propose the NEMMS which includes at least
one antiferromagnetic (AFM) layer (see Fig.~\ref{fig_nanorod})
that could be set into motion by spin-polarized current. Our idea
is based on the following facts: \emph{i}) theoretical predictions
\cite{xu:226602, Haney:2007(2), gomo:2010} and experimental
evidence \cite{Tsoi:2007, Urazhdin:2007, tang:122504, Dai08,
Tsoi-2008} of STT effects in AFMs; \emph{ii}) strong (compared to
FM) spin-lattice coupling in AFM that reveals itself, e.g., in the
pronounced magnetoelastic effects like an energy gap for AFMR
frequency  \cite{Borovik-Romanov:1984} and shape-induced magnetic
anisotropy \cite{Ryabchenko:2005,gomo:PhysRevB.75.174439}. In the
framework of hydrodynamic-like approach we analyze the coupled
magneto-mechanical dynamics of nanorod consisting of FM, NM and
AFM layers and calculate eigen frequencies and current-induced
mechanical and magnetic responses of the system. We show that
dissipative and nondissipative components of spin-polarized ac
current contribute differently to magneto-mechanical motion and
thus could be separated experimenttally.  The proposed device can
be also used as a current-driven nanoresonator that produces no
magnetic field.

The paper is devoted to the 80-th anniversary of the prominent
Ukrainian experimentalist Prof. V. V. Eremenko whose contribution
into the field of magnetoelasticity is remarkable and is
world-wide recognized.

\section{Model}\label{sec_model}
Let us consider the NEMMS that demonstrates the torsional
mechanical oscillations, e.g. doubly clamped nanorod
(Fig.~\ref{fig_nanorod} a). In general case, torsional dynamics
can be viewed as inhomogeneous (space-dependent) rotation of the
crystal lattice with respect to some reference state. On the other
hand, the magnetics with the strong enough exchange coupling
between the magnetic sublattices have another rotational degrees
of freedom, namely, those related with the solid-like rotation of
the magnetic sublattices \cite{Andreev:1980}. Lattice and magnetic
rotations could be coupled due to, e.g., magnetic anisotropy,
magnetoelastic or/and shape effects. Thus, any spin torque
transferred to the magnetic layer will induce twisting of the
crystal lattice and \emph{vise versa}, any mechanical torque will
induce rotations/oscillations of the magnetic subsystem.

In what follows we consider a  heterostructure that includes a
thin (thickness $d_{\mathrm{AFM}}$) metallic AFM layer inserted,
just in the middle between two metallic NM rods (each of the
length $L\gg d_{\mathrm{AFM}}$). Spin-polarized electric current
$J$ flowing through this system exerts spin torque to AFM layer
due to spin-flip processes at the NM/AFM interface. Thus, the
magnetic subsystem serves as a source of the magnetic and, as a
result, the mechanical torque for the whole system.

The optimal geometry of the magnetic (FM, or polarizer, and AFM,
or ``rotator'') layers can be predicted from general principles.
Curren-induced STT  is parallel to the FM magnetization,
$\mathbf{M}_{\mathrm{FM}}$, so, $\mathbf{M}_{\mathrm{FM}}$ should
be parallel to the axis of nanorod. On the other hand, the most
effective energy transfer between the magnetic and crystal
lattices occurs for the modes with the same symmetry. So, an
optimal orientation of  the magnetic vectors should allow
transversal (with respect to nanorod axis) oscillations with the
minimal possible frequency.

It should be noted that spin-polarized current acts on AFM layer
in three ways. First, STT that is proportional to the spin flux
transferred to the magnetic layer and is related with dissipative
processes. Second, spin current produces the effective magnetic
field $\mathbf{H}_{sd}\propto J \mathbf{M}_{\mathrm{FM}}$ parallel
to the spin polarization. Corresponding torque that acts on AFM
vector is nondissipative (adiabatic). Third, the current itself
generates an Oersted field which direction and value within an AFM
layer depends upon the geometry of the system. The last
contribution is supposed to produce a negligible effect on AFM
dynamics and will be disregarded in the following
consideration\footnote{According to Refs.
\cite{stiles-2007,Pufall:2007cond.mat2416P} typical value of
current-induced Oersted field  is 1 kOe. For FM materials with
characteristic fields of reorientation 0.1$\div 1$~kOe the effect
of Oersted field can be significant. However, in AFMs with strong
exchange coupling and high N\'{e}el temperature (FeMn, IrMn, NiO)
the typical value of spin-flop field is higher and falls into
1$\div 10$~kOe range. Thus, the effect of the Oersted field can be
neglected, at least in the first approximation.}. The value of the
effective field $\mathbf{H}_{sd}$ depends upon the exchange
coupling between free and localized spins (so called
$sd$-exchange) and thus can be noticeable, especially in the case
of ac current, as will be shown below.

Coupled rotational dynamics of the magnetic and crystal lattices
can be described phenomenologically in the framework of continuius
approach in terms of the Gibbs' vectors
$\boldsymbol{\varphi}_\alpha=\tan(\theta_\alpha/2)
\mathbf{e}_\alpha$ that parametrize solid-like rotation of the
crystal lattice ($\alpha\Rightarrow \mathrm{lat}$) and magnetic
subsystem ($\alpha\Rightarrow \mathrm{mag}$) around an
instantaneous rotation axis $\mathbf{e}_\alpha$ through the angle
$\theta_\alpha$. Vectors $\boldsymbol{\varphi}_\alpha(\mathbf{r},
t)$ are the field variables that define the state of the crystal
and magnetic lattices at a moment $t$ in a point $\mathbf{r}$. In
the simplest case under consideration (thin nanorod) the rotation
axis coinsides with the rod axis, so
$\mathbf{e}_\mathrm{lat}\|\mathbf{e}_\mathrm{mag}\|Z$.

Time, $\dot{\theta}_\alpha$, and space,
$\theta^\prime_\alpha\equiv\nabla_z\theta_\alpha$, derivatives of
thus introduced generalized coordinates $\theta_\mathrm{lat}$ and
$\theta_\mathrm{mag}$ generate the rotation frequencies and
vorticities, correspondingly\footnote{~In general case, frequency
is a vector and vorticity is a 2-nd rank tensor.}.

According to Ref. \cite{Andreev:1980}, the rotating magnetic frame
produces the dynamic contribution into macroscopic magnetization,
$M_{\mathrm{AFM}}$, of AFM. Thus, with account of the effective
magnetic field $\mathbf{H}_{sd}\|Z$  the magnetization of AFM
layer is parallel to the nanorod axis $Z$ and its value is
expressed as
\begin{equation}\label{eq-magnetization_AFM}
  M_{\mathrm{AFM}}=\frac{\chi}{\gamma}(\dot{\theta}_\mathrm{mag}+\gamma {H}_{sd})S_{\mathrm{AFM}}=\frac{\chi}{\gamma}(\dot{\theta}_\mathrm{mag}+\gamma \beta_{\mathrm{ad}} j)S_{\mathrm{AFM}},
\end{equation}
\noindent where $S_{\mathrm{AFM}}$ is the nanorod crossection area
within AFM layer,  $\chi$ is magnetic susceptibility, $\gamma$ is
gyromagnetic ratio. The last expression in
(\ref{eq-magnetization_AFM}) includes the material adiabatic (see
below) constant $\beta_{\mathrm{ad}}$ that defines the relation
between the effective field ${H}_{sd}=\beta_{\mathrm{ad}} j$ and
the
the current density $j=J/S_{\mathrm{AFM}}$\footnote{~Stricktly speaking, current density $j$ is defined by the effective (Sharvin) crossection which in the case of inhomogeneous rod can differ from $S_{\mathrm{AFM}}$.}. 
As follows from definition of the effective field
$\mathbf{H}_{sd}$, $\beta_{\mathrm{ad}}$ is proportional to the
constant of $sd$-exchange and to the fraction of free electrons
that did not flip their spins at NM/AFM interface. Thus, this
constant describes the action of nondissipative (\emph{ad}iabatic)
component of spin-polarized current, as will be discussed below.

 The Lagrange function of the system written from the general symmetry
considerations takes a form:
\begin{eqnarray}\label{eq_Lagrange_1}
  \mathcal{L}&=&\frac{1}{2}\int_{-L}^{L}dz\left[I(z)\dot{\theta}_\mathrm{lat}^2-\kappa(\theta^\prime_\mathrm{lat})^2\right]\\
 &+&S_{\mathrm{AFM}}\int_{-d_{\mathrm{AFM}}/2}^{d_{\mathrm{AFM}}/2}dz\left[\frac{\chi}{2\gamma^2}\left(\dot{\theta}_\mathrm{mag}+\gamma\beta_{\mathrm{ad}}
 j\right)^2-U(\theta_\mathrm{mag}-\theta_\mathrm{lat})\right].\nonumber
\end{eqnarray}
Here $\kappa$ is a torsion modulus (rigidity) that can be
expressed through the elastic modula and the dimensions of the
sample once the geometry is known,
$U(\theta_\mathrm{mag}-\theta_\mathrm{lat})$ is the energy of the
magnetic anisotropy which depends upon the relative orientation of
the magnetic moments with respect to crystal lattice (see
Fig.~\ref{fig_nanorod} b). A specific (per unit length) moment of
inertia  of nanorod, $I(z)\equiv \int
\rho_{\mathrm{rod}}(x^2+y^2)dx dy$, is supposed to be different in
NM, $I(z)\equiv I_{\mathrm{NM}}$, $d_{\mathrm{AFM}}/2\le |z|\le L$
and in AFM, $I(z)\equiv I_{\mathrm{AFM}}$, $|z|\le
d_{\mathrm{AFM}}/2$ regions, here $\rho_{\mathrm{rod}}$ is the
nanorod density. In Eq.~(\ref{eq_Lagrange_1}) we have neglected
inhomogeneous exchange interactions (terms with
$\theta^\prime_\mathrm{mag}$) that are vanishingly small for a
thin (below the characteristic domain wall thickness) AFM layer.
We also assume that $\kappa$ is constant along the rod,
generalization for a more complicated case is straightforward.

Dissipative phenomena within an AFM layer that arise from the STT
and internal damping are described with the help of generalized
potential (or Rayleigh dissipation function)
\cite{gomo:2011arXiv1106.4231G} as follows:
\begin{equation}\label{eq_Rayleigh_mag}
\mathcal{R}_{\mathrm{AFM}}=S_{\mathrm{AFM}}\int_{-d_{\mathrm{AFM}}/2}^{d_{\mathrm{AFM}}/2}dz\left(\chi\frac{\gamma_{\mathrm{AFM}}}{\gamma^2}\dot{\theta}_\mathrm{mag}^2-\frac{\beta_{\mathrm{dis}}j}{\gamma}
  \dot{\theta}_\mathrm{mag}\right),
\end{equation}
where  $\gamma_{\mathrm{AFM}}$ is a half-width of AFMR that
characterizes the damping. We have also taken into account that
the current polarization is parallel to the rod axis,
$\mathbf{M}_\mathrm{FM}\|Z$.

The above introduced  material constant $\beta_{\mathrm{dis}}$
that describes \emph{dis}sipative component of spin-polarized
current needs some special explanation. The value
$\beta_{\mathrm{dis}}j$ is equal to spin-flux that is transferred
to the unit volume of AFM layer due to spin-flip scattering of the
conduction electrons at NM/AFM interface. Thus, two constants,
$\beta_{\mathrm{ad}}$ and $\beta_{\mathrm{dis}}$, though having
different physical dimensions, are in a certain sense
complementary: the greater is one, the smaller is other.

Damping of the mechanical oscillations are accounted by the
corresponding Rayleigh function with the damping constant
$\gamma_{\mathrm{lat}}$:
\begin{equation}\label{eq_Rayleigh_el}
\mathcal{R}_{\mathrm{lat}}=\frac{1}{2}\int_{-L}^{L}dzI(z)\gamma_{\mathrm{lat}}\dot{\theta}_\mathrm{lat}^2.
\end{equation}

Functions (\ref{eq_Lagrange_1}), (\ref{eq_Rayleigh_mag}) and
(\ref{eq_Rayleigh_el}) together with the boundary conditions
$\theta_\mathrm{lat}(\pm L)=0$ (doubly clamped rod) generate the
system of dynamic equations for the angles $\theta_\mathrm{lat}$,
$\theta_\mathrm{mag}$ that unambiguously describes the nanorod
state. Oscillatory behavior of a system implies small deflections
of $\theta_\mathrm{lat}, \theta_\mathrm{mag}$ from equilibrium
zero values. To this end, magnetic anisotropy can be approximated
as $U(\theta_\mathrm{mag}-\theta_\mathrm{lat})\approx
\chi\Omega_\mathrm{AFMR}^2(\theta_\mathrm{mag}-\theta_\mathrm{lat})^2/(2\gamma^2)$,
where $\Omega_\mathrm{AFMR}$ is AFMR frequency of the mode that
corresponds to homogeneous\footnote{~As it was already mentioned
above, we consider only long-wave motions of AFM subsystem,
so-called macrospin approximation.} (within AFM layer) rotation of
the magnetic moments around $Z$-axis. It should be stressed that
the constant of magnetic anisotropy,
$K_\mathrm{AFM}\equiv\chi\Omega_\mathrm{AFMR}^2/\gamma^2$, is
defined by spin-orbit or dipole interactions and thus includes
contribution of magnetoelastic nature.

\section{Coupled magneto-mechanical dynamics}\label{sec_results}
Let us consider small oscillations induced by ac current
$j=j_0\cos\omega t$. Corresponding equations for the space
dependent functions $\theta_\mathrm{lat}(z)$ and
$\theta_\mathrm{mag}(z)$ in neglection of damping could be reduced
to a form:
\begin{eqnarray}\label{eq_dynamic_equation_1}
  &&\kappa\frac{d^2 \theta_\mathrm{lat}}{dz^2}+\omega^2\left[I(z)+\frac{S_{\mathrm{AFM}}\Theta(z)\chi\Omega_\mathrm{AFMR}^2}{\gamma^2(\Omega_\mathrm{AFMR}^2-\omega^2)}\right]\theta_\mathrm{lat}\nonumber\\
  &&=-S_{\mathrm{AFM}}\Theta(z)\frac{(\beta_{\mathrm{dis}}-i\chi\beta_{\mathrm{ad}}\omega)\Omega_\mathrm{AFMR}^2}{\gamma(\Omega_\mathrm{AFMR}^2-\omega^2)}j_0,\nonumber\\
&&\theta_\mathrm{mag}=\left[\frac{\Omega_\mathrm{AFMR}^2\theta_\mathrm{lat}}{\Omega_\mathrm{AFMR}^2-\omega^2}+\frac{\gamma(\beta_{\mathrm{dis}}-i\chi\beta_{\mathrm{ad}}\omega)}{\chi(\Omega_\mathrm{AFMR}^2-\omega^2)}j_0\right]\Theta(z),
\end{eqnarray}
where form-function $\Theta(z)=1$ inside the AFM layer ($|z|\le
d_{\mathrm{AFM}}/2$ ) and vanishes outside it ($|z|\ge
d_{\mathrm{AFM}}/2$)).

Analysis of Eqs.~(\ref{eq_dynamic_equation_1}) shows that the
spin-polarized current produces a mechanical torque (r.h.s. of the
first equation) and thus is a motive force for torsional
oscillations. The value of the torque is proportional to the
magnetic anisotropy constant
$K_\mathrm{AFM}\propto\Omega_\mathrm{AFMR}^2$ and the thickness of
AFM layer (factor $\Theta(z)$) and can increase greatly in the
vicinity of AFMR ($\omega\rightarrow\Omega_\mathrm{AFMR}$).
Physical interpretation of this fact is quite obvious: mechanical
torque occurs due to spin-lattice coupling within AFM layer and
should be proportional to its thickness and coupling constant, the
current acts directly on the magnetic subsystem and indirectly on
the mechanical one, thus the largest effect should be observed at
AFMR frequency.

\subsection{Oscillation modes and spectrum}
The rod under consideration has two types of the torsion eigen
modes, symmetric
($\theta_\mathrm{lat}(z)=\theta_\mathrm{lat}(-z)$) and
antisymmetric ($\theta_\mathrm{lat}(z)=-\theta_\mathrm{lat}(-z)$)
with respect to space inversion. From
Eqs.~(\ref{eq_dynamic_equation_1}) it follows that in the present
geometry the spin-polarized current can excite only symmetric
modes that show maximum deflection $\theta_\mathrm{lat}$ within an
AFM layer ($z\approx 0$).

\begin{figure}[htbp]
\begin{centering}
 \includegraphics[width=0.8\textwidth]{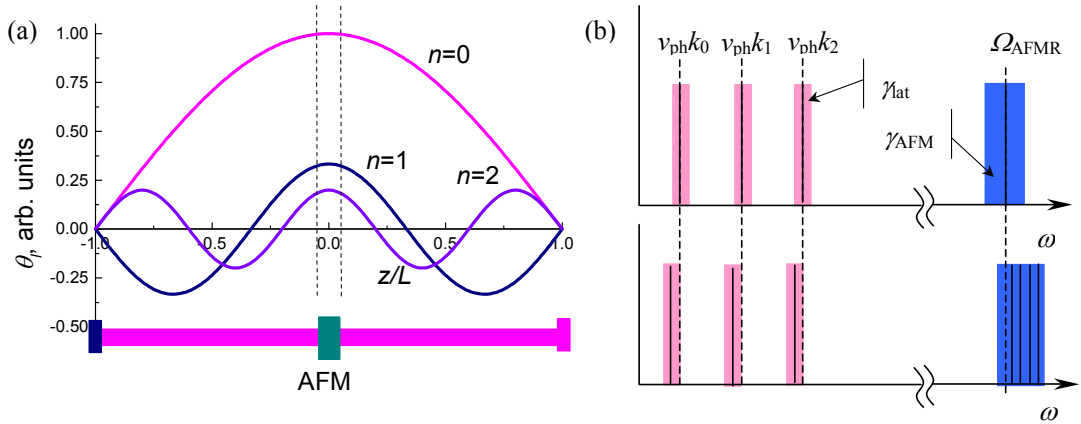}
  \caption{(Color online)\textbf{Torsional modes and spectrum of AFM-based
  nanorod}. (a) Low-frequency torsional modes,
  $\omega=v_{\mathrm{ph}}k_n$, $n=0,1,2$ induced by STT. Relative amplitude of torsional angle, $\theta_\mathrm{lat}(z)$,
  is frequency dependent. Low panel schematically shows the position of AFM layer (the thickness $t_{\mathrm{AFM}}=0.02L$ is slightly exaggerated).
  (b) Spectrum of eigen modes (scematically). In the absence of coupling (upper
  panel) the mechanical modes though smeared (half-width
  $\gamma_{\mathrm{lat}}$) are well separated due to the rather high
  value of quality factor $Q_\mathrm{lat}$. The magnetic modes ($\omega=\Omega_\mathrm{AFMR}$) are
  degenerated and have a pronounced width
  ($\gamma_{\mathrm{AFM}}$). Magneto-mechanical coupling (lower
  panel) results in the ``red'' shift of the mechanical modes and
  small ``blue'' shift of the magnetic modes (shown by solid
  vertical lines). While the shifted mechanical modes are still
  well distinguishable, the spectrum of the shifted magnetic modes falls completely
  into the line width. \label{fig_spectrum}}
 \end{centering}
\end{figure}

In the first approximation (taking into account that
$d_{\mathrm{AFM}}/L\ll 1$) the symmetric modes (see
Fig.~\ref{fig_spectrum} a)  could be represented as
\begin{eqnarray}\label{eq_eigen_modes_mec}
  \theta^{(n)}_\mathrm{lat}(z,t)&=&\theta^{(n)}_\mathrm{lat}(0)e^{i\omega^{(n)}t}\cos k_nz,\\
  \theta^{(n)}_\mathrm{mag}(z,t)&=&\frac{\Omega_\mathrm{AFMR}^2\Theta(z)}{\Omega_\mathrm{AFMR}^2-\omega^2}\theta^{(n)}_\mathrm{lat}(0)e^{i\omega^{(n)}t},\nonumber
\end{eqnarray}
were the allowed wave vector $k_n=\pi(2n+1)/(2L)$ is calculated
from the boundary conditions. Corresponding eigen frequencies
$\omega^{(n)}$
 calculated from
Eqs.~(\ref{eq_dynamic_equation_1}) are the following:
\begin{eqnarray}\label{eq_eigen_frequencies}
    \omega^{(n)}_\pm&=&\frac{1}{\sqrt{2}}\left\{\Omega_\mathrm{AFMR}^2+(1+\lambda_n)v_{\mathrm{ph}}^2k_n^2\phantom{\frac{1}{{2}}}\right.\nonumber\\
    &\pm&\left.\left[\left(\Omega_\mathrm{AFMR}^2-(1+\lambda_n)v_{\mathrm{ph}}^2k_n^2\right)^2+4\lambda_nv_{\mathrm{ph}}^2k_n^2\Omega_\mathrm{AFMR}^2\right]^{1/2}\right\}^{1/2},
\end{eqnarray}
where $v_{\mathrm{ph}}=(\kappa/\overline{I})^{1/2}$ is the phonon
velocity and $\overline{I}\equiv (1/2L)\int_{-L}^{L}I(z)dz$ is the
averaged moment of inertia. Following the notions of
Ref.~\cite{Kovalev:2006}, we have introduced in
Eq.~(\ref{eq_eigen_frequencies})  the coupling coefficient
\begin{equation}\label{eq_coupling_coef}
  \lambda_n\equiv\frac{K_{\mathrm{AFM}}V_{\mathrm{AFM}}}{2L\overline{I}v_{\mathrm{ph}}^2k_n^2}
\end{equation}
which is proportional to the magnetic anisotropy of the whole AFM
layer (with the volume $V_{\mathrm{AFM}}\equiv
d_{\mathrm{AFM}}S_{\mathrm{AFM}}$). Expression
(\ref{eq_eigen_frequencies}) for eigen frequencies is analogous to
one obtained in Ref.~\cite{Kovalev:2006} for a nanorod with the FM
layer.

The expression (\ref{eq_eigen_frequencies}) confirms quite obvious
conclusion that the spectrum of nanorod consists of two branches
--- high-frequency quasimagnetic, $\omega^{(n)}_+$, and
low-frequency quasimechanical (torsional), $\omega^{(n)}_-$. In
the limit $\lambda_n\rightarrow 0$ the quasimagnetic frequency
$\omega^{(n)}_+\rightarrow\Omega_\mathrm{AFMR}$ and
quasimechanical one $\omega^{(n)}_-\rightarrow
v_{\mathrm{ph}}k_n$.

Further analysis of current-induced dynamics can be simplified due
to specification of ``small'' and ``large'' quantities. The
frequency of the torsional fundamental, ``zero'', mode for a
nanosized rod ($L\propto 30\div 100$~nm, $v_{\mathrm{ph}}\propto
5\cdot10^3$~m/s) is $v_{\mathrm{ph}}k_0\propto 10\div 100$~GHz.
Characteristic AFMR frequency for a bulk sample of a typical AFM
with high N\'{e}el temperature (FeMn, IrMn, NiO) is noticeably
greater, $\nu_\mathrm{AFMR}\equiv \Omega_\mathrm{AFMR}/2\pi\propto
150\div 1000$~GHz\footnote{~For the small samples
$\nu_\mathrm{AFMR}$ can be smaller due to the size effects, see,
e.g.\cite{Mishra:2010}.}, depending on the mode type
\cite{Endoh:1973, nishitani:221906,
Ivanov:PhysRevLett.105.077402}. So, in contrast to FM, where the
fundamental frequency of the mechanical oscillations is close to
the FMR frequency \cite{Kovalev:2006}, for the nanorods with AFM
layer $\Omega_\mathrm{AFMR}\gg v_{\mathrm{ph}}k_0$. However, for
higher harmonics (with $n\propto 10\div 100$) the crossing of
frequencies ($v_{\mathrm{ph}}k_n\propto\Omega_\mathrm{AFMR}$) is
possible.

The coupling constants
$\lambda_n<\lambda_{n-1}<\ldots<\lambda_0\ll 1$. For example, for
a typical AFM Ir$_{20}$Mn$_{80}$ the anisotropy constant
$K_{\mathrm{AFM}} \propto 10^5$~J/m$^3$ \cite{Mishra:2010}, so,
for the $50\times50\times2$~nm AFM layer $\lambda_0\propto
10^{-2}$. However, it should be stressed that the constant
$\lambda_0$ in AFM is substantially larger than for analogous FM
layer (e.g., for Fe the value $\lambda_0\propto 10^{-3}$
\cite{Kovalev:2006}), due to the difference in magnetic
anisotropy.

The quality factor of the mechanical oscillations,
$Q_\mathrm{lat}=v_{\mathrm{ph}}k_0/(2\gamma_{\mathrm{lat}})$,
strongly depends upon the surface effects but even in the worst
case is as large as $10^3$ \cite{davis:072513}. The quality factor
of the metallic magnetic subsystem,
$Q_\mathrm{mag}=\Omega_\mathrm{AFMR}/(2\gamma_{\mathrm{AFM}})$, is
much smaller, e.g., for the metallic FM the quality factor
$Q_\mathrm{mag}\propto 10^2$ \cite{Kovalev:2006}.

Thus, the spectrum of the mechanical and magnetic excitations
(Eq.~(\ref{eq_eigen_frequencies})) for a typical AFM-based nanorod
has the following features (see Fig.~\ref{fig_spectrum}b):
\renewcommand{\theenumi}{\roman{enumi}}
\renewcommand{\labelenumi}{\emph{\theenumi})}
\begin{enumerate}
  \item in the absence of coupling ($\lambda=0$) the spectrum of the mechanical
  modes consists of thin ($Q_\mathrm{lat}\gg 1$) well-separated lines. The
  spectrum of the magnetic modes is degenerated
  ($\omega=\Omega_\mathrm{AFMR}$), corresponding line is rather thick;
    \item far from the crossing the coupling-induced shift of the frequencies,
    $\omega^{(n)}_-=v_{\mathrm{ph}}k_n(1-\lambda_nv_{\mathrm{ph}}^2k_n^2/2\Omega_\mathrm{AFMR}^2)$, $\omega^{(n)}_+=\Omega_\mathrm{AFMR}(1+\lambda_nv_{\mathrm{ph}}^2k_n^2/2\Omega_\mathrm{AFMR}^2)$, is
    vanishingly small. So, ``mechanical'' modes are still well separated, while the splitting of the ``magnetic'' modes is below the line width;
  \item in the vicinity of crossing the splitting of the mechanical
  and magnetic modes is substantially greater,
  $\omega^{(n)}_\pm=\Omega_\mathrm{AFMR}(1\pm\sqrt{\lambda_n}/2)$. Damping processes are defined mainly by the magnetic subsystem, so, corresponding quality
  factor is close to $Q_\mathrm{mag}$. Thus, the magnetic and mechanical modes
  could be resolved providing $\sqrt{\lambda_n}Q_\mathrm{mag}>1$.
\end{enumerate}
\subsection{Current-induced oscillations}
From the properties of oscillation spectrum it follows that
current-induced behavior of nanorod is different in the
low-frequency ($\omega\ll\Omega_\mathrm{AFMR}$) and high-frequency
($\omega\propto\Omega_\mathrm{AFMR}$) ranges. Let us consider them
separately.

In the low-frequency range the last term in the l.h.s. of the first
of Eqs.~(\ref{eq_dynamic_equation_1}) is small ($\propto \lambda$)
and can be neglected. To this end, torsion angle of mechanical
oscillations is expressed as
\begin{equation}\label{eq_low_f_theta_l}
    \theta_\mathrm{lat}(z;\omega)=\frac{
V_{\mathrm{AFM}}j_0}{\gamma\overline{I}L}\frac{\pi}{4\omega
v_{\mathrm{ph}}k_0}\frac{(\beta_{\mathrm{dis}}-i\chi\beta_{\mathrm{ad}}\omega)\sin[(L-|z|)\omega/c]}{\sqrt{\cos^2(L\omega/c)+(\pi/4Q_\mathrm{lat})^{2}\sin^2(L\omega/c)}}e^{i\phi},
\end{equation}
where $\phi$ is the frequency dependent phase shift with respect
to $j$, in the vicinity of resonance $\phi\rightarrow \pi/2$.

It can be easily seen from Eq.~(\ref{eq_low_f_theta_l}) that the
current-induced torsional oscillations have clearly defined
resonance character at $\omega=\omega_{-}^{(n)}\approx
v_{\mathrm{ph}}k_n$. Space dependence of $\theta_\mathrm{lat}(z)$
at a given $\omega$ (see Fig.~\ref{fig_spectrum} a) is close to
the mechanical eigen modes. The resonant amplitude obtained from
Eq.~(\ref{eq_low_f_theta_l}) is
\begin{eqnarray}\label{eq_low_f_theta_l_amp}
    \theta_\mathrm{lat}^{(n)}(\mathrm{res})&=&\frac{Q_\mathrm{lat}
V_{\mathrm{AFM}}j_0}{\gamma\overline{I}Lv_{\mathrm{ph}}^2k_0^2}\left(\frac{i\beta_{\mathrm{dis}}}{2n+1}+\chi\beta_{\mathrm{ad}}v_{\mathrm{ph}}k_0\right)\nonumber\\
&=&\frac{2\lambda_0Q_\mathrm{lat} \gamma
j_0}{\Omega_\mathrm{AFMR}^2\chi}\left(\frac{i\beta_{\mathrm{dis}}}{2n+1}+\chi\beta_{\mathrm{ad}}v_{\mathrm{ph}}k_0\right).
\end{eqnarray}
Here the factor $i$ reflects the phase shift of the torsion angle
with respect to current.

As seen from Eq.~(\ref{eq_low_f_theta_l_amp}), rotation of lattice
results from two effects induced by spin-polarized current,
namely, dissipative STT ($\propto \beta_{\mathrm{dis}}$) and
adiabatic effective spin-induced field ($\propto
\beta_{\mathrm{ad}}$). The first contribution diminishes with the
frequency ($\propto n$) growth, while the second one is frequency
independent (at least, for $\omega\ll\Omega_\mathrm{AFMR}$).
Moreover, STT-induced term is phase-shifted with respect to
current, while adiabatic term is in phase with current. This opens
a way to separate these contributions by measuring current
dependence of resonant torsional oscillations.

An amplitude of the corresponding magnetic oscillations differs
from $\theta_\mathrm{lat}^{(n)}(\mathrm{res})$ by the factor
$(1+2i\lambda_0Q_\mathrm{lat})$, as seen from the following
\begin{equation}\label{eq_low_f_theta_m}
    \theta_\mathrm{mag}^{(n)}(\mathrm{res})=\frac{\gamma
j_0}{\chi\Omega_\mathrm{AFMR}^2}(1+2i\lambda_0Q_\mathrm{lat})\left(\frac{\beta_{\mathrm{dis}}}{2n+1}-i\chi\beta_{\mathrm{ad}}v_{\mathrm{ph}}k_0\right).
\end{equation}
It also depends upon both dissipative and nondissipative
current-induced contributions, however, phase shift with respect
to current is much more complicated due to the term with
$\lambda_0Q_\mathrm{lat}$. Time derivative
$\dot{\theta}_\mathrm{mag}^{(n)}(\mathrm{res})=iv_{\mathrm{ph}}k_n\theta_\mathrm{mag}^{(n)}(\mathrm{res})$
is proportional to magnetization of AFM layer (see
Eq.~(\ref{eq-magnetization_AFM}) and thus can be detected
experimentally.

In the high-frequency range the magnetic modes with different $n$
are almost degenerated. So, the current induces mechanical,
\begin{equation}\label{eq_high_f_theta_l_amp}
     \theta_\mathrm{lat}(\mathrm{res})=\frac{15Q_\mathrm{AFM}
V_{\mathrm{AFM}}j_0}{16\gamma\overline{I}L\Omega_\mathrm{AFMR}^2}\left(i\beta_{\mathrm{dis}}+\chi\beta_{\mathrm{ad}}\Omega_\mathrm{AFMR}\right),
\end{equation}
and magnetic,
\begin{equation}\label{eq_high_f_theta_m}
     \theta_\mathrm{mag}(\mathrm{res})=-\frac{\gamma Q_\mathrm{AFM}
j_0}{\chi\Omega_\mathrm{AFMR}^2}\left(i\beta_{\mathrm{dis}}+\chi\beta_{\mathrm{ad}}\Omega_\mathrm{AFMR}\right)\left(1+\frac{15v_{\mathrm{ph}}^2k_0^2
}{8\Omega_\mathrm{AFMR}^2}\lambda_0Q_\mathrm{AFM}\right)
\end{equation}
oscillations with the frequency
$\omega\approx\Omega_\mathrm{AFMR}$.
\section{Conclusions}\label{sec_consclusions}
In the present paper we considered new aspect of magneto-elastic
interactions and studied magnetomechanical oscillations induced by
spin-polarized current for the simplest of twisting nanorod. Our
calculations demonstrate that ac spin-polarized current can excite
quasimechanical (torsional) as well as quasimagnetical modes.

It is interesting to note that the ac spin-polarized current
affects the AFM layer in the case of strong scattering at NM/AFM
interface (due to STT effect) and in the case of weak scattering
as well (due to the effective $sd$-exchange field ``injected''
with free electrons into AFM layer). Ratio between dissipative and
nondissipative contribution is proportional to the phase shift
between mechanical oscillations and current and thus can be
measured experimentally in the low frequency range.

An amplitude of quasimechanical mode depends upon the geometry of
the sample (see Eq.~(\ref{eq_low_f_theta_l_amp})) and can be
enhanced by diminishing the moment of intertia (e.g. by using
carbon nanotubes \cite{Williams:PhysRevLett.89.255502}) and by
enlarging AFM volume $V_\mathrm{AFM}$. However, if the thickness
of AFM layer, $d_\mathrm{AFM}$, becomes greater than the free path
of spin-polarized electrons, contribution of dissipative (STT)
part will be reduced.

The effectiveness  of the described
electric-through-magnetic-to-mechanical energy conversion can be
increased by using nanorod with periodical FM/NM/AFM structure,
however this system needs additional treatment and is out of scope
of this paper.

In this work we considered torsional oscillations of the
effectively one dimensional structure. Analogous  results could be
obtained for nanobeams that show flexional oscillations.

The authors acknowledge partial  financial support from the
Special Program for Fundamental Research of the Department of
Physics and Astronomy of National Academy of Sciences of Ukraine.
The work of H.G. and S.K. was partially supported by the grant
from the Ministry of Education and Science of Ukraine.
%

 \end{document}